\documentclass[usenatbib,usegraphicx]{mn2e}
\usepackage{graphicx}
\usepackage{natbib}
\usepackage[dvipdfm]{hyperref}
\usepackage{amssymb}
\usepackage{bm}
\usepackage{color}
\usepackage[switch]{lineno}

\newcommand {\be} {\begin{equation}}
\newcommand {\ee} {\end{equation}}
\newcommand {\bea} {\begin{eqnarray}}
\newcommand {\eea} {\end{eqnarray}}

\newcommand{\eg}{{\it e.g.}}

\title[Energetics of a VHE Flare in PKS~1222+216]{Energetic Constraints on a Rapid Gamma-Ray Flare in PKS~1222+216}
\author[Nalewajko et~al.]
{Krzysztof Nalewajko,$^1$\thanks{E-mail: knalew@jila.colorado.edu}
Mitchell C. Begelman,$^{1,2}$
Beno{\^i}t Cerutti,$^3$
\newauthor
Dmitri A. Uzdensky$^3$
and Marek Sikora$^4$\\
$^1$JILA, University of Colorado and National Institute of Standards and Technology, 440 UCB, Boulder, CO 80309, USA\\
$^2$Department of Astrophysical and Planetary Sciences, University of Colorado, UCB 389, Boulder, CO 80309, USA\\
$^3$Center for Integrated Plasma Studies, Department of Physics, University of Colorado, UCB 390, Boulder, CO 80309, USA\\
$^4$Nicolaus Copernicus Astronomical Centre, Bartycka 18, 00-716 Warsaw, Poland}

\begin{document}

\maketitle

\begin{abstract}
We study theoretical implications of a rapid Very-High-Energy (VHE) flare detected by MAGIC in the Flat-Spectrum Radio Quasar PKS~1222+216. The minimum distance from the jet origin at which this flare could be produced is $0.5\;{\rm pc}$. A moderate Doppler factor of the VHE source, $\mathcal{D}_{\rm VHE}\sim 20$, is allowed by all opacity constraints. The concurrent High-Energy (HE) emission observed by {\it Fermi} provides estimates of the total jet power and the jet magnetic field strength. Energetic constraints for the VHE flare are extremely tight: for an isotropic particle distribution they require a huge co-moving energy density in the emitting region and a very efficient radiative process. We disfavor hadronic processes due to their low radiative efficiency, as well as the synchrotron scenario recently proposed for the case of HE flares in the Crab Nebula, since the parameters needed to overcome the radiative losses are quite extreme. The VHE emission can be explained by the Synchrotron Self-Compton (SSC) mechanism for $\mathcal{D}_{\rm VHE}\sim 20$ or by the External Radiation Compton (ERC) mechanism involving the infrared radiation of the dusty torus for $\mathcal{D}_{\rm VHE}\sim 50$. After discussing several alternative scenarios, we propose that the extreme energy density constraint can be satisfied when the emission comes from highly anisotropic short-lived bunches of particles formed by the kinetic beaming mechanism in magnetic reconnection sites. By focusing the emitting particles into very narrow beams, this mechanism allows one to relax the causality constraint on the source size, decreasing the required energy density by 4 orders of magnitude.
\end{abstract}


\section{Introduction}

Relativistic jets are responsible for bright emission of blazars that occasionally shows violent variability on a wide range of time scales. The brightest and most rapid flares place the tightest constraints on the source energetics, radiative efficiency and geometry. Fast variability of blazars can be probed by telescopes of the highest effective area, i.e. Imaging Atmospheric Cerenkov Telescopes (IACTs), which are sensitive to Very-High-Energy (VHE) photons, $\sim 0.1-10\;{\rm TeV}$. Two BL Lacertae objects (BL Lacs) were shown to exhibit a variability time scale of a few minutes at an apparent luminosity of the order of $10^{46}\;{\rm erg\,s^{-1}}$: PKS~2155-304, observed by H.E.S.S. in July 2006 \citep{2007ApJ...664L..71A,2009A&A...502..749A,2010A&A...520A..83H}, and Mrk~501, observed by MAGIC in June and July 2005 \citep{2007ApJ...669..862A}. These flares appear to be produced in emitting regions smaller than the characteristic size of supermassive black holes producing the jets. This source compactness poses several challenges, including a strong constraint on the minimum energy density and a very high Doppler factor in order to avoid the absorption of VHE gamma rays \citep{2008MNRAS.384L..19B}.

Equally surprising were detections of a handful of blazars belonging to the subclass of Flat-Spectrum Radio Quasars (FSRQs) at VHE energies, primarily because their VHE emission is expected to be absorbed internally, if produced within the Broad-Line Region (BLR) with characteristic radius of $\sim 0.1-0.3\;{\rm pc}$.
3C~279 was detected by MAGIC in early 2006 \citep{2008Sci...320.1752M} and, at redshift 0.536, remains the most distant known VHE source. PKS~1510-089 was observed by H.E.S.S. in March 2010 \citep{Wagner10}, while PKS~1222+216 was detected by MAGIC in June 2010 \citep{2011ApJ...730L...8A}. The last observation is particularly interesting, because the VHE flux is variable on the time scale of $\simeq 10$ minutes. Thus, the case of PKS~1222+216 combines the difficulties posed by rapid flux variability and a strong radiative environment. This is the first evidence that extremely fast VHE variability in blazars can originate at the parsec scale.

\cite{2011ApJ...730L...8A} noted that the simultaneous requirements for a very small source radius and its relatively distant location from the jet origin pose a serious challenge. They briefly proposed several alternative solutions, including
localized emitting sites or very efficient jet recollimation. \cite{2011A&A...534A..86T} studied the origin of the VHE emission in greater detail, in particular by fitting a two-zone emission model to the broad-band Spectral Energy Distribution (SED) of PKS~1222+216. They adopted a very high Doppler factor of 75 for the compact `blob' producing the VHE flare, but rejected the possibility that it could result from bulk acceleration in a magnetic reconnection site in the `minijets' scenario \citep{2009MNRAS.395L..29G,2011MNRAS.413..333N} because of low expected jet magnetization at parsec scales. With a very high Doppler factor, they strongly favored the External Radiation Compton (ERC) mechanism over Synchrotron Self-Compton (SSC) in their interpretation of the VHE emission. They concluded that the most likely explanation of the rapid VHE flare in PKS~1222+216 is extreme jet focusing due to a recollimation shock in the scenario proposed originally by \cite{2009ApJ...699.1274B}.

In this work, we investigate the energetic constraints on the source of the VHE flare in PKS~1222+216 and examine the ability of various radiative processes to explain this phenomenon. In Section \ref{sec_obs}, we explore the observational constraints, calculate contributions of different radiation components to the gamma-ray opacity, and estimate the jet magnetic field strength. In Section \ref{sec_ene}, we derive the energetic constraints that need to be satisfied regardless of the radiative mechanism involved. In Section \ref{sec_rad}, we compare the radiative efficiencies of several processes: ERC, SSC, electron synchrotron, proton synchrotron and photo-meson. In Section \ref{sec_dis}, we discuss the constraints on the jet power, Doppler factor and degree of collimation efficiency, and evaluate possible interpretations of this flare. We propose that the source energetics can be explained by a high level of anisotropy and inhomogeneity recently found to be associated with magnetic reconnection sites \citep{2012arXiv1205.3210C}. Our results are summarized in Section \ref{sec_sum}.

The Lorentz factor is $\Gamma=(1-\beta^2)^{-1/2}$ and the Doppler factor is $\mathcal{D}=[\Gamma(1-\beta\cos\theta_{\rm obs})]^{-1}$, where $\beta=v/c$ is the dimensionless velocity and $\theta_{\rm obs}$ is the viewing angle. We distinguish between the jet Lorentz factor $\Gamma_{\rm j}$ and the VHE emission source Lorentz factor $\Gamma_{\rm VHE}$, as well as between the corresponding Doppler factors. We denote quantities measured in the jet co-moving frame with $'$, and those measured in the VHE emitter co-moving frame with $''$. We will refer to Doppler factors satisfying $\mathcal{D}_{\rm VHE}\simeq\Gamma_{\rm VHE}\simeq\Gamma_{\rm j}\sim 20$ as ``typical'' and those of $\mathcal{D}_{\rm VHE}\simeq\Gamma_{\rm VHE}\gg 20$ as ``high''. Particle energies are denoted by $\mathcal{E}$, and electric field strengths by $E$. Symbols with a numerical subscript should be read as a dimensionless number $X_{\rm n}=X/(10^n\,{\rm cgs\,units})$. We adopt the standard cosmology with $H_0=71\;{\rm km\,s^{-1}\,Mpc^{-1}}$, $\Omega_{\rm m}=0.27$ and $\Omega_{\rm \Lambda}=0.73$.

\section{Observational constraints}
\label{sec_obs}

PKS~1222+216 (4C~+21.35) is located at redshift $z=0.432$, which corresponds to a luminosity distance of $d_{\rm L}=2.4\;{\rm Gpc}=7.3\times 10^{27}\;{\rm cm}$. The IACT system MAGIC detected this source on 2010 June 17 (MJD 55364.92) during a 30-min observation \citep{2011ApJ...730L...8A}. After correction for absorption by the Extragalactic Background Light (EBL), its spectrum was fitted in the energy range between $\mathcal{E}_{\rm VHE,min,obs}=70\;{\rm GeV}$ and $\mathcal{E}_{\rm VHE,max,obs}=400\;{\rm GeV}$ (we adopt a typical photon energy of $\mathcal{E}_{\rm VHE,obs}\simeq 100\;{\rm GeV}$) with a power-law model $N(\mathcal{E})\propto \mathcal{E}^{-\Gamma_{\rm VHE}}$ with photon index of $\Gamma_{\rm VHE}=2.7\pm 0.3$ and integrated flux of $F_{\rm VHE,obs}=(2.3\pm 0.5)\times 10^{-10}\;{\rm erg\,s^{-1}\,cm^{-2}}$ (based on their Figure 3), which corresponds to an isotropic luminosity of $L_{\rm VHE}=4\pi d_{\rm L}^2F_{\rm VHE,obs}\simeq 1.5\times 10^{47}\;{\rm erg\,s^{-1}}$. They were also able to calculate a light curve using 6-min time bins. The first 4 data points show a strong flux increase, with the flux doubling time scale of $t_{\rm VHE,obs}\simeq 10\;{\rm min}=600\;{\rm s}$. The last data point indicates a decay time scale of the same order of magnitude.

Concurrently with the VHE flare, a strong High-Energy (HE; $\sim 0.1-10\;{\rm GeV}$) flare was observed by the {\it Fermi} Large Area Telescope (LAT). The luminosity of the HE flare was $L_{\rm HE}\simeq 10^{48}\;{\rm erg\,s^{-1}}$ and the observed variability time scale was crudely estimated at $t_{\rm HE,obs}\simeq 1\;{\rm d}$ \citep{2011ApJ...733...19T}. A clear spectral break was observed at the energy of $\simeq 2\;{\rm GeV}$ (with integration time of 8 days), similar to many other bright FSRQs. Such breaks may result from absorption of gamma rays by the ionized helium Ly$\alpha$ continuum \citep{2010ApJ...717L.118P}\footnote{For alternative models, see, e.g., \cite{2010ApJ...714L.303F,2012MNRAS.420...84Z}.}. This would be an indication that the HE radiation is produced within the BLR. \cite{2011arXiv1110.4471F} presented an independent analysis of the same data, looking for hints of even faster variability. They identified a subflare peaking at MJD 55365.63, $17\;{\rm h}$ after the MAGIC observation, and estimated the flux rising time scale at $\simeq 6\;{\rm h}$. They also calculated a SED with a 1-day integration time, which indicates a possible turnover around $\simeq 20\;{\rm GeV}$.

Optical flux observed in 2010 showed daily variability and moderate linear polarization of $\simeq 5\%$, with little correspondence to the gamma-ray flux variations \citep{2011arXiv1110.6040S}. This indicates that a non-thermal synchrotron component contributes to the optical flux. VLBI monitoring at 7 mm wavelength (43 GHz) revealed a superluminal component ejected around February/March 2010 (MJD $\simeq 55260$) with apparent propagation velocity of $\simeq 14c$ \citep{2012IJMPS..08..356J}. These authors also report a concurrent rotation of the optical polarization angle by $\simeq 200^\circ$.

\begin{figure}
\includegraphics[width=\columnwidth]{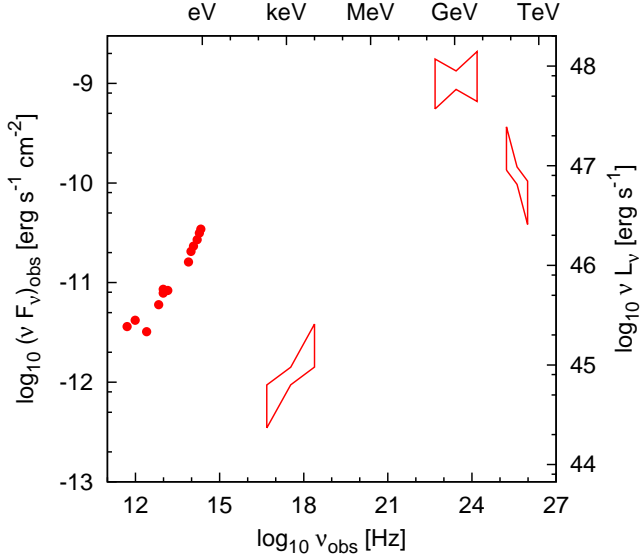}
\caption{Spectral energy distribution (SED) of PKS~1222+216, including (from high to low energies) MAGIC spectrum corrected for absorption by Extragalactic Background Light, quasi-simultaneous {\it Fermi}/LAT, {\it Swift}/XRT and {\it Swift}/UVOT data, and {\it Spitzer} data from Malmrose et~al. (2011). Adapted from Figure 1 in Tavecchio et~al. (2011).}
\label{fig_sed}
\end{figure}

The low-energy SED of PKS~1222+216 has been analyzed by \cite{2011A&A...534A..86T} (Figure \ref{fig_sed}). With the low level of the synchrotron emission, two thermal components can be clearly identified. The optical/UV spectrum is very hard and appears to be dominated by thermal accretion disk emission of luminosity $L_{\rm d}\simeq 5\times 10^{46}\;{\rm erg\,s^{-1}}$, in addition to the weak synchrotron component. The IR spectrum indicates the presence of a dusty torus of luminosity $L_{\rm IR}\simeq 10^{46}\;{\rm erg\,s^{-1}}$, hence its covering factor is $\xi_{\rm IR} = L_{\rm IR}/L_{\rm d}\simeq 0.2$ \citep{2011ApJ...732..116M}. \cite{2011A&A...534A..86T} also estimate the covering factor of the broad-line region at $\xi_{\rm BLR}\simeq 0.02$, hence its luminosity at $L_{\rm BLR} = \xi_{\rm BLR}L_{\rm d}\simeq 10^{45}\;{\rm erg\,s^{-1}}$. The BLR and the dusty torus form the radiative environment of the jet. From the general knowledge of their properties, we can estimate the energy density profiles of their emission along the jet \citep[see][for a review]{2009ApJ...704...38S}. The characteristic BLR radius is
\be
r_{\rm BLR}\simeq 0.1\;{\rm pc}\times L_{\rm d,46}^{1/2}\simeq 0.22\;{\rm pc}\simeq 7\times 10^{17}\;{\rm cm}
\ee
\citep{2008MNRAS.386..945T}, while the inner radius of the dusty torus for the sublimation temperature of $T_{\rm IR}\simeq 1200\;{\rm K}$ is
\be
r_{\rm IR}\simeq 4\;{\rm pc}\times L_{\rm d,46}^{1/2}T_{\rm IR,3}^{-2.6}\simeq 5.6\;{\rm pc}\simeq 1.7\times 10^{19}\;{\rm cm}
\ee
\citep{2008ApJ...685..160N}\footnote{It is possible that the inner radius of the torus is somewhat smaller than this estimate \citep{2011ApJ...732..116M}, but this would not change our results.}. Within these inner radii, the energy densities in the external frame are roughly uniform and equal to
\bea
u_{\rm BLR}(r<r_{\rm BLR}) &\simeq& \frac{L_{\rm BLR}}{4\pi r_{\rm BLR}^2c}\simeq 6\times 10^{-3}\;{\rm erg\,cm^{-3}}\,;\\
u_{\rm IR}(r<r_{\rm IR}) &\simeq& \frac{L_{\rm IR}}{4\pi r_{\rm IR}^2c}\simeq 9\times 10^{-5}\;{\rm erg\,cm^{-3}}\,.
\eea
The typical energy of photons emitted from the BLR is $\mathcal{E}_{\rm BLR}\simeq \mathcal{E}_{\rm Ly\alpha}\simeq 10\;{\rm eV}$ and that of the IR photons $\mathcal{E}_{\rm IR}\simeq 3k_{\rm B}T_{\rm IR}\simeq 0.3\;{\rm eV}$.

Blazar jets have opening angles that on average satisfy the relation $\theta_{\rm j}\sim 0.2/\Gamma_{\rm j}$ \citep{2009A&A...507L..33P}. We use it to estimate the typical jet radius at a given distance $r$ from the supermassive black hole:
\be
R_{\rm j}'(r)\simeq \frac{r}{5\Gamma_{\rm j}}\simeq 1.5\times 10^{16}\;{\rm cm}\times \left(\frac{r}{r_{\rm min}}\right)\left(\frac{\Gamma_{\rm j}}{20}\right)^{-1}\,,
\ee
where $r_{\rm min}\simeq 0.5\;{\rm pc}$ is the minimum distance required to avoid the absorption of gamma rays by the broad emission lines (see Section \ref{sec_obs_opac}). The extremely short variability time scale of the VHE flare introduces a very tight constraint on the size of the VHE emitting region. Its maximum radius and opening angle are
\bea
R_{\rm VHE}'' &\simeq& \frac{\mathcal{D}_{\rm VHE}ct_{\rm VHE,obs}}{(1+z)}\simeq 2.5\times 10^{14}\;{\rm cm}\times\left(\frac{\mathcal{D}_{\rm VHE}}{20}\right),\\
\label{eq_theta_vhe}
\theta_{\rm VHE} &\simeq& \frac{R_{\rm VHE}''}{r}\simeq 1.7\times 10^{-4}\left(\frac{\mathcal{D}_{\rm VHE}}{20}\right)\left(\frac{r}{r_{\rm min}}\right)^{-1}\,,
\eea
respectively. The variability time scale of the HE flare constrains the size of the HE emitting region to
\be
R_{\rm HE}' \simeq \frac{\mathcal{D}_{\rm j}ct_{\rm HE,obs}}{(1+z)}\simeq 4\times 10^{16}\;{\rm cm}\times\left(\frac{\mathcal{D}_{\rm j}}{20}\right)\,,
\ee
which corresponds to jet distance $r_{\rm HE}\simeq 5\Gamma_{\rm j}R_{\rm HE}'\simeq 1.2\;{\rm pc}\times(\mathcal{D}_{\rm j}/20)^2(\mathcal{D}_{\rm j}/\Gamma_{\rm j})^{-1}$.

\subsection{Opacity for gamma rays}
\label{sec_obs_opac}

The external radiation can absorb gamma-rays produced within the jet in the photon-photon pair production process. This process operates between an absorbed (``hard'') photon of energy $\mathcal{E}_{\rm hard}$ and an absorbing (``soft'') photon of energy $\mathcal{E}_{\rm soft}$ when $\mathcal{E}_{\rm hard}\mathcal{E}_{\rm soft}/(m_{\rm e}c^2)^2>1$. The peak cross section for this process, $\sigma_{\rm \gamma\gamma}\simeq 0.2\sigma_{\rm T}$, where $\sigma_{\rm T}$ is the Thomson cross section, is achieved for $\mathcal{E}_{\rm hard}\mathcal{E}_{\rm soft}/(m_{\rm e}c^2)^2\simeq 3.6$. The IR radiation can absorb hard photons of observed energy
\be
\mathcal{E}_{\rm hard,IR,obs}\gtrsim \frac{(m_{\rm e}c^2)^2}{(1+z)\mathcal{E}_{\rm IR}}\simeq 600\;{\rm GeV}\,,
\ee
which is beyond the energy range of the MAGIC observation. For the BLR radiation, the observed threshold is
\be
\label{eq_Ehard_BLR}
\mathcal{E}_{\rm hard,BLR,obs}\gtrsim \frac{(m_{\rm e}c^2)^2}{(1+z)\mathcal{E}_{\rm BLR}}\simeq 18\;{\rm GeV}\,.
\ee
The optical depth to gamma-ray photons can be as high as $\tau_{\rm \gamma\gamma,BLR}=\sigma_{\rm \gamma\gamma}n_{\rm BLR}r_{\rm BLR}\simeq 32$, where $n_{\rm BLR}=u_{\rm BLR}/\mathcal{E}_{\rm BLR}$ is the number density of the BLR photons within $r_{\rm BLR}$. Thus, if the gamma-ray emission is produced within the BLR, we would expect strong absorption features in the observed SED. However, in PKS~1222+216 there is no evidence for absorption by hydrogen Ly$\alpha$ photons between the MAGIC spectrum and the quasi-simultaneous {\it Fermi}/LAT spectrum \citep{2011ApJ...730L...8A}. The minimum distance $r_{\rm min}$ of the VHE source from the central black hole can be estimated from the dependence of the Ly$\alpha$ absorption threshold on the scattering angle. The minimum scattering angle is given by \citep[\eg,][]{1967PhRv..155.1404G}
\be
\chi_{\rm min}=\arccos\left[1-\frac{2(m_{\rm e}c^2)^2}{(1+z)\mathcal{E}_{\rm VHE,max,obs}\mathcal{E}_{\rm BLR}}\right]\simeq 25^\circ\,.
\ee
For the idealized case of a flat BLR contained within the accretion plane, which minimizes $r_{\rm min}$, and the jet coincident with the BLR symmetry axis, we can simply write $r_{\rm min}\simeq r_{\rm BLR}/\tan\chi_{\rm min}\simeq 0.5\;{\rm pc}$. This estimate should be increased if the BLR radial emissivity distribution has a significant `tail' for $r>r_{\rm BLR}$, or if the BLR is extended vertically. Nevertheless, we can safely state that the VHE emission detected by MAGIC is produced at least at the parsec scale.

We also consider absorption of gamma rays by the soft radiation co-produced within the VHE source. This soft radiation would be observed at energies
\be
\label{eq_Esoft_vhe}
\mathcal{E}_{\rm soft,VHE,obs}\simeq \frac{3.6(\mathcal{D}_{\rm VHE}m_{\rm e}c^2)^2}{(1+z)^2\mathcal{E}_{\rm VHE,obs}}\simeq 1.8\;{\rm keV}\times \left(\frac{\mathcal{D}_{\rm VHE}}{20}\right)^2.
\ee
For $\mathcal{D}_{\rm VHE}\simeq \mathcal{D}_{\rm j}$, this is obviously the same result as in the previous case. The optical depth inflicted by this internal radiation is
\bea
\label{eq_tauVHE}
\tau_{\rm \gamma\gamma,VHE}&\simeq& \sigma_{\rm \gamma\gamma}n_{\rm soft}''R_{\rm VHE}''\simeq\nonumber\\
&\simeq& \frac{\sigma_{\rm T}L_{\rm soft}}{20\pi c(1+z)\mathcal{D}_{\rm VHE}^3R_{\rm VHE}''\mathcal{E}_{\rm soft,VHE,obs}}\simeq\nonumber\\
&\simeq& 0.04\times L_{\rm soft,45}\left(\frac{\mathcal{D}_{\rm VHE}}{20}\right)^{-6}\,.
\eea
Here, we use the co-moving energy density valid for a moving source:
\be
u_{\rm soft}''\simeq \frac{L_{\rm soft}''}{4\pi cR_{\rm VHE}''^2}\simeq \frac{L_{\rm soft}}{4\pi c\mathcal{D}_{\rm VHE}^4R_{\rm VHE}''^2}\,.
\ee
If we instead chose a transformation for a stationary source, as in the previous case, the resulting optical depth would increase by a factor of $\simeq 3\ln{(2\Gamma_{\rm VHE})}$, which value is $\simeq 11$ for $\Gamma_{\rm VHE}=20$, and $\simeq 14$ for $\Gamma_{\rm VHE}=50$. Hence, the optical depth is below unity even if the VHE source is stationary. The fact that the internal absorption of the VHE emission is insignificant for $\mathcal{D}_{\rm VHE}\simeq 20$ appears to be in conflict with the work of \cite{2008MNRAS.384L..19B}. We discuss this issue in Section \ref{sec_dis_lec_dop}.

\subsection{Magnetic field strength}
\label{sec_obs_mag}

The magnetic field strength can be estimated by modeling the broad-band SED of quasars in the External Radiation Compton (ERC) scenario (see Section \ref{sec_rad_erc}). The Compton dominance parameter is defined as the luminosity ratio of the ERC and synchrotron components, $q=L_{\rm ERC}/L_{\rm SYN}$. When the Compton scattering is in the Thomson regime, we have $q\simeq u_{\rm ext}'/u_{\rm B}'$, where $u_{\rm ext}'$ is the energy density of external radiation, while $u_{\rm B}'=B'^2/(8\pi)$ is the magnetic energy density. For PKS~1222+216, we can use the HE emission, which dominates the overall SED, and the optical emission, which is produced mainly by the thermal component from the accretion disk with a small contribution from the synchrotron component. From the SED compiled by \cite{2011A&A...534A..86T}, we estimate the Compton dominance parameter at $q\simeq 100$. At the distance where the HE emission is most likely produced, we have
\be
\label{eq_Bj}
B_{\rm j}'(r_{\rm HE}) \simeq \Gamma_{\rm j}\left[\frac{8\pi u_{\rm BLR}(r_{\rm HE})}{q}\right]^{1/2}\simeq \frac{0.14\;{\rm G}}{q_{\rm 2}^{1/2}}\left(\frac{\mathcal{D}_{\rm j}}{20}\right)^{-1}\,,
\ee
where we assumed that the energy density of the broad emission lines scales like $u_{\rm BLR}(r)\propto r^{-2}$ for $r>r_{\rm BLR}$. If the magnetic field is dominated by the toroidal component and the jet does not accelerate, which is generally thought to be the case at the parsec scale, the magnetic field strength should depend on the location in the jet as $B_{\rm j}'\propto r^{-1}$. Hence, the magnetic field strength at a distance of $1\;{\rm pc}$ is $B_{\rm j}'(1\;{\rm pc})\simeq 0.17\;{\rm G}$. The Poynting flux corresponding to this value is
\bea
\label{eq_LB}
L_{\rm B} &\simeq& (\pi R_{\rm j}'^2)(\Gamma_{\rm j}^2u_{\rm B,j}')c \simeq\nonumber\\
&\simeq& 10^{45}\;{\rm erg\,s^{-1}}\times\left(\frac{B_{\rm j}'\,r}{0.17\;{\rm G\,pc}}\right)^2\left(\frac{r}{5\Gamma_{\rm j}R_{\rm j}'}\right)^{-2}.
\eea

A more direct method for estimating the magnetic field strength is based on VLBI measurements of the shift in the absolute position of the radio core as a function of the observing frequency. Typical results at a distance of $1\;{\rm pc}$ are $0.1-0.3\;{\rm G}$ \citep[\eg,][]{2009MNRAS.400...26O}. Although we know of no such results for PKS~1222+216, these values are consistent with the one that we calculated for PKS~1222+216 from the Compton dominance.

We note that \cite{2011A&A...534A..86T} adopted a different jet Lorentz factor of $\Gamma_{\rm j}\simeq 10$ and a distance of the HE emitting region that corresponds to $B_{\rm j}'(1\;{\rm pc})\simeq 0.035\;{\rm G}$, which is a factor of $\simeq 5$ lower than the value adopted in this work.

\section{Energetic constraints}
\label{sec_ene}

In this section, we discuss energetic constraints on a generic radiative process involving ultra-relativistic particles, that need to be satisfied in order to explain the VHE flare in PKS~1222+216. Let $\mathcal{E}_{\rm 1part}''$ be the particle energy required to produce a spectral component peaking at $\mathcal{E}_{\rm VHE,obs}$, and $L_{\rm 1part,em}''$ be the emission luminosity of a single such particle, both measured in the co-moving frame of the VHE source. The co-moving cooling time scale is $t_{\rm cool}''=\mathcal{E}_{\rm 1part}''/L_{\rm 1part,em}''$, and the cooling ratio can be defined as $Q=t_{\rm VHE}''/t_{\rm cool}''=R_{\rm VHE}''L_{\rm 1part,em}''/(\mathcal{E}_{\rm 1part}''c)$, where $t_{\rm VHE}''=R_{\rm VHE}''/c$ is the co-moving light-crossing time scale of the emitting region. The total energy carried by the emitting particles should not exceed the total energy that can be supplied through the region boundaries $\dot{\mathcal{E}}_{\rm in}''\simeq u_{\rm j}''(\beta_{\rm in}c)A_{\rm VHE}''$, where $\beta_{\rm in}$ is the dimensionless inflow velocity and $A_{\rm VHE}''$ is the effective co-moving surface area of the emitting region boundary. The energy dissipation rate related to the VHE emission is $\dot{\mathcal{E}}_{\rm diss}''=\eta_{\rm diss,VHE}\dot{\mathcal{E}}_{\rm in}''$. The relation between $L_{\rm VHE}''$ and $\dot{\mathcal{E}}_{\rm diss}''$ also depends on the cooling ratio. For $Q\gg 1$, all of the supplied energy can be immediately radiated away and thus we expect $L_{\rm VHE}''\simeq\dot{\mathcal{E}}_{\rm diss}''$. For $Q\ll 1$, only the $Q$ fraction of the supplied energy can be radiated away, hence $L_{\rm VHE}''\simeq Q\dot{\mathcal{E}}_{\rm diss}''$. Effectively, we write $L_{\rm VHE}''\simeq {\rm min}\{Q,1\}\dot{\mathcal{E}}_{\rm diss}''$, from which we find an estimate of the jet energy density:
\be
\label{eq_ujvhe2}
u_{\rm j,VHE}''\simeq \frac{1}{{\rm min}\{Q,1\}}\left(\frac{L_{\rm VHE}''}{\eta_{\rm diss,VHE}\beta_{\rm in}c A_{\rm VHE}''}\right)\,.
\ee
A typical value for the average dissipation efficiency in blazars is $\eta_{\rm diss,VHE}\simeq 0.1$ \citep[\eg,][]{2008MNRAS.385..283C}.

We will now relate the total co-moving luminosity $L_{\rm VHE}''$ to the luminosity observed in the external frame $L_{\rm VHE}$. The apparent luminosity in the external frame may be affected by two effects: a relativistic boost by a factor of $\mathcal{D}_{\rm VHE}^4$, and a geometric boost, due to the particle anisotropy in the co-moving frame, by a factor of $4\pi/\Omega_{\rm e}''$, where $\Omega_{\rm e}''$ is the solid angle covered by the particle beam. Hence, we write $L_{\rm VHE}\simeq (4\pi/\Omega_{\rm e}'')\mathcal{D}_{\rm VHE}^4L_{\rm VHE}''$. Using this and adopting a spherical geometry, we can calculate the energy density within the VHE source as
\bea
\label{eq_LEC_uj}
u_{\rm j,VHE}'' &\simeq& \frac{1}{\eta_{\rm diss,VHE}\eta_{\rm rad}}\left(\frac{\Omega_{\rm e}''}{4\pi}\right)\left(\frac{3L_{\rm VHE}}{4\pi c\mathcal{D}_{\rm VHE}^4R_{\rm VHE}''^2}\right)\simeq\nonumber\\
&\simeq& \frac{1200\;{\rm erg\,cm^{-3}}}{\eta_{\rm diss,VHE,-1}\eta_{\rm rad}}\left(\frac{\Omega_{\rm e}''}{4\pi}\right)\left(\frac{\mathcal{D}_{\rm VHE}}{20}\right)^{-6}\,.
\eea
We call this relation the \emph{Local Energetic Constraint} (LEC). If the jet energy density is homogeneous across the jet radius $R_{\rm j}'$, we can relate $u_{\rm j,VHE}''$ to the total jet power:
\bea
\label{eq_LEC_Lj}
L_{\rm j,VHE,LEC} &\simeq& (\pi R_{\rm j}'^2)(\Gamma_{\rm VHE}^2u_{\rm j,VHE}'')c \simeq\nonumber\\
&\simeq& \frac{3L_{\rm VHE}}{4\eta_{\rm diss,VHE}\eta_{\rm rad}}\left(\frac{\Omega_{\rm e}''}{4\pi}\right)\left(\frac{\Gamma_{\rm VHE}^2}{\mathcal{D}_{\rm VHE}^4}\right)\times\nonumber\\
&&\times\left(\frac{R_{\rm j}'}{R_{\rm VHE}''}\right)^2\simeq\frac{10^{49}\;{\rm erg\,s^{-1}}}{\eta_{\rm diss,VHE,-1}\eta_{\rm rad}}\times\nonumber\\
&&\times\left(\frac{\Omega_{\rm e}''}{4\pi}\right)\left(\frac{\mathcal{D}_{\rm VHE}}{20}\right)^{-6}\left(\frac{\Gamma_{\rm VHE}}{\Gamma_{\rm j}}\right)^2\times\nonumber\\
&&\times\left(\frac{r}{5\Gamma_{\rm j}R_{\rm j}'}\right)^{-2}\left(\frac{r}{r_{\rm min}}\right)^2\,.
\eea
This value is certainly too high to be realistic (see Section \ref{sec_dis_lec_pow}). There are several ways in which it can be lowered: by increasing the Doppler factor $\mathcal{D}_{\rm VHE}$, by focusing the emitting particles to a small $\Omega_{\rm e}''$, or by decreasing the jet radius $R_{\rm j}'$. One can also abandon the underlying assumption that the jet is homogeneous across its entire radius.

In the limit of $R_{\rm j}'\to R_{\rm VHE}''$, Equation (\ref{eq_LEC_Lj}) reduces to what we call the \emph{Global Energetic Constraint} (GEC):
\bea
\label{eq_GEC_vhe}
L_{\rm j,VHE,GEC} &\simeq& \frac{3L_{\rm VHE}}{4\eta_{\rm diss,VHE}\eta_{\rm rad}}\left(\frac{\Omega_{\rm e}''}{4\pi}\right)\left(\frac{\Gamma_{\rm VHE}^2}{\mathcal{D}_{\rm VHE}^4}\right)\simeq\nonumber\\
&\simeq& \frac{2.8\times 10^{45}\;{\rm erg\,s^{-1}}}{\eta_{\rm diss,VHE,-1}\eta_{\rm rad}}\left(\frac{\Omega_{\rm e}''}{4\pi}\right)\times\nonumber\\
&&\times\left(\frac{\mathcal{D}_{\rm VHE}}{20}\right)^{-2}\left(\frac{\mathcal{D}_{\rm VHE}}{\Gamma_{\rm VHE}}\right)^{-2}.
\eea
It can be interpreted either as the total jet power required in the case of extreme recollimation or as the fraction of the total jet power associated with the VHE source. It provides a solid lower limit on the total jet. However, since $L_{\rm VHE}\ll L_{\rm HE}$, the GEC is even more constraining for the HE flare. We can use an analogous formula to that in Equation (\ref{eq_GEC_vhe}):
\bea
\label{eq_GEC_he}
L_{\rm j,HE} &\simeq& \frac{3L_{\rm HE}}{4\eta_{\rm diss,HE}\eta_{\rm rad}}\left(\frac{\Omega_{\rm e}'}{4\pi}\right)\left(\frac{\Gamma_{\rm j}^2}{\mathcal{D}_{\rm j}^4}\right)\simeq\\
&\simeq& \frac{1.9\times 10^{46}\;{\rm erg\,s^{-1}}}{\eta_{\rm diss,HE,-1}\eta_{\rm rad}}\left(\frac{\Omega_{\rm e}'}{4\pi}\right)\left(\frac{\mathcal{D}_{\rm j}}{20}\right)^{-2}\left(\frac{\mathcal{D}_{\rm j}}{\Gamma_{\rm j}}\right)^{-2}.\nonumber
\eea
The jet Lorentz factor $\Gamma_{\rm j}$ and the dissipation efficiency $\eta_{\rm diss}$ are unlikely to be significantly higher than 20 and 0.1, respectively, while the particles producing the HE emission in the ERC process are likely to be less anisotropic than the more energetic ones producing the VHE emission. Hence, we accept this value as the best estimate of the total jet power in PKS~1222+216 during the concurrent HE-VHE flares \citep[see also][]{2011ApJ...733...19T}. The co-moving jet energy density associated with the HE flare is
\bea
\label{eq_uj_he}
u_{\rm j,HE}' &\simeq& \frac{1}{\eta_{\rm diss,HE}\eta_{\rm rad}}\left(\frac{\Omega_{\rm e}'}{4\pi}\right)\left(\frac{3L_{\rm HE}}{4\pi c\mathcal{D}_{\rm j}^4R_{\rm HE}'^2}\right)\simeq\nonumber\\
&\simeq& \frac{0.4\;{\rm erg\,cm^{-3}}}{\eta_{\rm diss,HE,-1}\eta_{\rm rad}}\left(\frac{\Omega_{\rm e}'}{4\pi}\right)\left(\frac{\mathcal{D}_{\rm j}}{20}\right)^{-6}\,.
\eea
The fact that $u_{\rm j,HE}'\ll u_{\rm j,VHE}''$, unless the Doppler factor of the VHE source is extremely high, is consistent with the large difference between the jet power estimates $L_{\rm j,HE}$ and $L_{\rm j,VHE,LEC}$.

Since the LEC is extremely tight, it is clear that a high radiative efficiency, with $Q\gtrsim 1$, is critical for the VHE flare energetics. Also, in the case of $Q\ll 1$, VHE emitting particles could spread far outside the VHE source, substantially increasing the observed variability time scale. As we will show in the next section, the $Q\gtrsim 1$ requirement for typical jet parameters is quite demanding.

\section{Efficiency of radiative processes}
\label{sec_rad}

Here, we consider various radiative processes in the context of the VHE flare in PKS~1222+216: Inverse Compton (IC) scattering (Section \ref{sec_rad_ic}), in particular External Radiation Compton (ERC, Section \ref{sec_rad_erc}) and Synchrotron Self-Compton (SSC, Section \ref{sec_rad_ssc}); synchrotron radiation (Section \ref{sec_rad_syn}); and hadronic processes (Section \ref{sec_rad_hadr}). For each process, we calculate the cooling ratio and provide additional constraints on parameters of the emitting region. These results are compared and summarized in Section \ref{sec_rad_comp}.

\subsection{Inverse Compton (IC) scattering}
\label{sec_rad_ic}

Inverse Compton scattering of soft radiation fields off ultra-relativistic electrons is the most successful model of gamma-ray emission in blazars and many other astrophysical sources. The co-moving energy of electrons that can upscatter soft photons of energy $\mathcal{E}_{\rm soft}''$ to produce the VHE emission is $\gamma_{\rm e,IC}\simeq [(1+z)\mathcal{E}_{\rm VHE,obs}/(\mathcal{D}_{\rm VHE}\mathcal{E}_{\rm soft}'')]^{1/2}$. The scattering proceeds in the Thomson regime if $b_{\rm soft}=\gamma_{\rm e,IC}\mathcal{E}_{\rm soft}''/(m_{\rm e}c^2)<1$, which translates to
\be
\mathcal{E}_{\rm soft}''<\frac{\mathcal{D}_{\rm VHE}(m_{\rm e}c^2)^2}{(1+z)\mathcal{E}_{\rm VHE,obs}}\simeq 36\;{\rm eV}\times\left(\frac{\mathcal{D}_{\rm VHE}}{20}\right)\,.
\ee
In such a case, the luminosity of a single electron is $L_{\rm 1e,IC}''\simeq \sigma_{\rm T}cu_{\rm soft}''\gamma_{\rm e,IC}^2$, where $u_{\rm soft}''$ is the energy density of soft photons, and the cooling ratio is
\bea
Q_{\rm IC} &\simeq& b_{\rm soft}\sigma_{\rm T}n_{\rm soft}''R_{\rm VHE}''\simeq\nonumber\\
&\simeq& 17\left(\frac{\mathcal{D}_{\rm VHE}}{20}\right)^{1/2}\left(\frac{\mathcal{E}_{\rm soft}''}{1\;{\rm eV}}\right)^{-1/2}u_{\rm soft,0}''\,,
\eea
where $n_{\rm soft}''=u_{\rm soft}''/\mathcal{E}_{\rm soft}''$ is the number density of soft photons.

In the context of FSRQs, the most relevant sources of soft radiation for IC scattering are various forms of external radiation (Section \ref{sec_rad_erc}) and the local synchrotron radiation (Section \ref{sec_rad_ssc}).

\subsubsection{External Radiation Compton (ERC)}
\label{sec_rad_erc}

Energy density of the external radiation in the co-moving frame is boosted by a factor of $u_{\rm ext}''/u_{\rm ext}\simeq \Gamma_{\rm VHE}^2$, while the photon energy increases by a factor of $\mathcal{E}_{\rm ext}''/\mathcal{E}_{\rm ext}\simeq \Gamma_{\rm VHE}$. Hence, the Thomson limit is $\mathcal{E}_{\rm ext}<1.8\;{\rm eV}\times (\mathcal{D}_{\rm VHE}/\Gamma_{\rm VHE})$ and the cooling ratio for the ERC process is
\bea
Q_{\rm ERC} &\simeq& 1500\times u_{\rm ext,0}\left(\frac{\mathcal{D}_{\rm VHE}}{20}\right)^2\left(\frac{\mathcal{D}_{\rm VHE}}{\Gamma_{\rm VHE}}\right)^{-3/2}\times\nonumber\\
&&\times\left(\frac{\mathcal{E}_{\rm ext}}{1\;{\rm eV}}\right)^{-1/2}.
\eea

The main components of external radiation in FSRQs are the broad emission lines and the thermal infrared emission from the dusty torus. Since the gamma-ray opacity arguments constrain the VHE emission to be produced outside the BLR, here we do not consider the ERC(BLR) process, but only the ERC(IR). The IR radiation, with a typical photon energy of $\mathcal{E}_{\rm IR}\simeq 0.3\;{\rm eV}$, is scattered in the Thomson regime. The electron Lorentz factor required for the ERC(IR) scenario is
\be
\gamma_{\rm e,ERC(IR)}\simeq 3.5\times 10^4\times\left(\frac{\mathcal{D}_{\rm VHE}}{20}\right)^{-1}\left(\frac{\mathcal{D}_{\rm VHE}}{\Gamma_{\rm VHE}}\right)^{1/2}\,.
\ee
For a VHE source located within $r_{\rm IR}$, the cooling ratio is
\be
Q_{\rm ERC(IR)}\simeq 0.25\left(\frac{\mathcal{D}_{\rm VHE}}{20}\right)^2\left(\frac{\mathcal{D}_{\rm VHE}}{\Gamma_{\rm VHE}}\right)^{-3/2}\,,
\ee
which is slightly below unity for a typical Doppler factor, but exceeds unity for $\mathcal{D}_{\rm VHE}\gtrsim 40$.

\subsubsection{Synchrotron Self-Compton (SSC)}
\label{sec_rad_ssc}

The Lorentz factor required to produce SSC emission peaking at VHE energies is
\bea
\gamma_{\rm e,SSC} &\simeq& \left[\frac{(1+z)\mathcal{E}_{\rm VHE,obs}}{20\,{\rm neV}\times\mathcal{D}_{\rm VHE}B_0''}\right]^{1/4}\simeq\nonumber\\
&\simeq& 2.4\times 10^4\times B_0''^{-1/4}\left(\frac{\mathcal{D}_{\rm VHE}}{20}\right)^{-1/4}\,.
\eea
Here, $B_0''=B''/(1\;{\rm G})$, while the term $20\;{\rm neV}=0.274\;{\rm G}\times he/(m_{\rm e}c)$ is the characteristic synchrotron photon energy for $B''=1\;{\rm G}$ and in the limit of $\gamma_{\rm e}\to 1$ (`neV' stands for `nanoelectron-volt'). The observed energy of the soft synchrotron radiation is
\bea
\mathcal{E}_{\rm SYN,obs} &\simeq& 20\,{\rm neV}\frac{\mathcal{D}_{\rm VHE}B_0''\gamma_{\rm e,SSC}^2}{1+z}\simeq\nonumber\\
&\simeq& 0.17\;{\rm keV}\times B_0''^{1/2}\left(\frac{\mathcal{D}_{\rm VHE}}{20}\right)^{1/2}.
\eea
The scattering proceeds in the Thomson regime for
\bea
B'' &<& \frac{(m_{\rm e}c^2)^4}{20\,{\rm neV\,G^{-1}}}\left[\frac{\mathcal{D}_{\rm VHE}}{(1+z)\mathcal{E}_{\rm VHE,obs}}\right]^3 \simeq\nonumber\\
&\simeq& 9\,{\rm G}\times\left(\frac{\mathcal{D}_{\rm VHE}}{20}\right)^3,
\eea
which is easily satisfied for parsec-scale jets (see Section \ref{sec_obs_mag}). The co-moving energy density of the soft synchrotron radiation is calculated from the relation $L_{\rm VHE}=L_{\rm SSC}\simeq (u_{\rm SYN}''/u_{\rm B}'')L_{\rm SYN}$:
\bea
u_{\rm SYN}'' &\simeq& \frac{B''}{4\pi\mathcal{D}_{\rm VHE}^2R_{\rm VHE}''}\left(\frac{L_{\rm VHE}}{2c}\right)^{1/2}\simeq\nonumber\\
&\simeq& 1.3\;{\rm erg\,cm^{-3}}\times B_0''\left(\frac{\mathcal{D}_{\rm VHE}}{20}\right)^{-3}.
\eea
The cooling ratio of the SSC process is
\bea
Q_{\rm SSC} &\simeq& \frac{\sigma_{\rm T}R_{\rm VHE}''u_{\rm SYN}''\gamma_{\rm e,SSC}}{m_{\rm e}c^2}\simeq \nonumber\\
&\simeq& 6.3\times B_0''^{3/4}\left(\frac{\mathcal{D}_{\rm VHE}}{20}\right)^{-9/4}.
\eea
For magnetic fields in the range of $B''\simeq 0.1-1\;{\rm G}$ and for typical Doppler factors, this cooling ratio is about one order of magnitude higher than for the ERC(IR) process. However, in contrast to the ERC processes, it strongly decreases with increasing Doppler factor.

\subsection{Synchrotron radiation}
\label{sec_rad_syn}

Synchrotron emission from FSRQ blazars usually peaks in the IR band and does not extend beyond the UV band. This corresponds to maximum electron Lorentz factors of $\gamma_{\rm e}\simeq 10^{3-4}$. The same electrons will also produce an ERC component extending to GeV energies. Electrons of much higher energies will scatter the external radiation very inefficiently, because the scattering will proceed deeply in the Klein-Nishina regime. But their synchrotron emission may extend into the HE band, the evidence for which we find in the flares recently detected in the Crab Nebula \citep{2011Sci...331..739A,2011ApJ...741L...5S,2012ApJ...749...26B}. If electrons can be accelerated to such extreme energies in blazars, their synchrotron components could extend even beyond the HE band, in part due to the relativistic Doppler effect. Here, we consider the possibility that the VHE emission from PKS~1222+216 is of synchrotron origin.

In order to produce emission observed at VHE energies, we require electrons of Lorentz factor
\bea
\label{eq_gammae_syn}
\gamma_{\rm e,SYN} &\simeq& \left[\frac{(1+z)\mathcal{E}_{\rm VHE,obs}}{20\,{\rm neV}\times\mathcal{D}_{\rm VHE}B_0''}\right]^{1/2}\simeq\nonumber\\
&\simeq& 6\times 10^8\times B_0''^{-1/2}\left(\frac{\mathcal{D}_{\rm VHE}}{20}\right)^{-1/2}\,.
\eea
The luminosity of a single electron is $L_{\rm 1e,SYN}''\simeq \sigma_{\rm T}cu_{\rm B}''\gamma_{\rm e,SYN}^2$. The cooling ratio for the synchrotron process is
\bea
\label{eq_Qsyn}
Q_{\rm SYN} &\simeq& \frac{\sigma_{\rm T}u_{\rm B}''R_{\rm VHE}''\gamma_{\rm e,SYN}}{m_{\rm e}c^2}\simeq\nonumber\\
&\simeq& 4900\times B_0''^{3/2}\left(\frac{\mathcal{D}_{\rm VHE}}{20}\right)^{1/2}\,.
\eea
The requirement that $Q_{\rm SYN}\gtrsim 1$ can be satisfied already for $B''>3\;{\rm mG}\times (\mathcal{D}_{\rm VHE}/20)^{-1/3}$. For magnetic field strengths typical of parsec-scale jets, the cooling ratio for the synchrotron process can be 3 orders of magnitude higher than that for the ERC(IR) process.

\subsubsection{Extreme electron acceleration}
\label{sec_rad_syn_acc}

Although the cooling ratio for the synchrotron process in producing the VHE emission is very high, strong radiative losses pose a problem for accelerating electrons to the required energy \citep[\eg,][]{1996ApJ...457..253D,2010MNRAS.405.1809L,2011ApJ...737L..40U}. If the acceleration proceeds in a uniform electric field $E''$ and a uniform magnetic field $B''$, then the maximum electron Lorentz factor allowed by the radiative losses is $\gamma_{\rm e,max}\simeq (6\pi eE''/\sigma_{\rm T})^{1/2}/B_\perp''$, where $B_\perp''$ is the magnetic field component perpendicular to the electric field. The observed energy of the corresponding synchrotron radiation is
\bea
\mathcal{E}_{\rm SYN} &\simeq& 20\;{\rm neV\,G^{-1}}\times\frac{6\pi e\mathcal{D}_{\rm VHE}}{(1+z)\sigma_{\rm T}}\left(\frac{E''}{B_\perp''}\right)\simeq\nonumber\\
&\simeq& 4\;{\rm GeV}\times\left(\frac{\mathcal{D}_{\rm VHE}}{20}\right)\left(\frac{E''}{B_\perp''}\right)\,.
\eea
In order to explain the VHE emission of PKS~1222+216 with synchrotron radiation, we require $E''/B_\perp''\simeq 26(\mathcal{D}_{\rm VHE}/20)^{-1}$. This condition violates the ideal MHD approximation, $\bm{E}''\simeq \bm{B}''\times\bm\beta$. However, it can be satisfied in magnetic reconnection sites, where the MHD limit does not apply. \cite{2004PhRvL..92r1101K} described a mechanism of extreme particle acceleration in magnetic reconnection layers, in which sufficiently energetic electrons follow relativistic Speiser orbits, which tend to focus them into the region located around the layer midplane, where $B_\perp''$ vanishes, and $E''/B_\perp''>1$ locally. This mechanism has been applied successfully to the case of the Crab Nebula flares \citep{2011ApJ...737L..40U,2012ApJ...746..148C}. Still, the requirement that $E''/B_\perp''\simeq 26$ is rather severe, as it can only be satisfied in a tiny fraction of the reconnection layer volume. In the scenario applied to the Crab Nebula flares, $E''/B_\perp''\simeq 4$ was sufficient. We would need an extremely high Doppler factor of $\mathcal{D}_{\rm VHE}\simeq 130$ in order to have the same situation in the case of PKS~1222+216.

In a magnetic reconnection layer, the electric field $E_{\rm acc}''$ that accelerates electrons is induced by the reconnecting magnetic field $B_{\rm VHE}''$. Therefore, we have $E_{\rm acc}''\simeq\beta_{\rm rec}B_{\rm VHE}''$, where $\beta_{\rm rec}$ is the dimensionless reconnection rate, i.e. the inflow velocity of the reconnecting plasma. Even in the relativistic regime, the reconnection rate is limited to $\beta_{\rm rec}\lesssim 0.1$ \citep{2005MNRAS.358..113L}. The magnetic energy cannot be stored within the reconnection layer; instead it needs to be continuously supplied. Thus, the Local Energetic Constraint (LEC) for the case of magnetic reconnection has to be calculated by following the energy inflow approach. Hence, we take Equation (\ref{eq_ujvhe2}) and substitute $u_{\rm B,VHE}''$ for $u_{\rm j,VHE}''$, and $\beta_{\rm rec}$ for $\beta_{\rm in}$. The effective surface area of a flat double-faced reconnection layer is $A_{\rm VHE}''\simeq 2\pi R_{\rm VHE}''^2$. We obtain the following constraint on the jet magnetic field strength:
\bea
\label{eq_Bvhe_rec}
B_{\rm VHE}'' &\gtrsim& \left(\frac{\Omega_{\rm e}''L_{\rm VHE}}{\eta_{\rm diss,VHE}\eta_{\rm rad}\pi\beta_{\rm rec}c\mathcal{D}_{\rm VHE}^4R_{\rm VHE}''^2}\right)^{1/2}\simeq\nonumber\\
&\simeq& \frac{130\;{\rm G}}{\eta_{\rm rad}^{1/2}}\Omega_{\rm e}''^{1/2}\eta_{\rm diss,VHE,-1}^{-1/2}\beta_{\rm rec,-1}^{-1/2}\times\nonumber\\
&&\times\left(\frac{\mathcal{D}_{\rm VHE}}{20}\right)^{-3}\,.
\eea
This value is about 3 orders of magnitude higher than what we expect in a parsec-scale jet (see Section \ref{sec_obs_mag}). It can be lowered by one order of magnitude either by increasing the Doppler factor to $\mathcal{D}_{\rm VHE}\simeq 43$ or by decreasing the electron beam solid angle to $\Omega_{\rm e}''\simeq 0.01\;{\rm sr}$. For example, in the context of the Crab Nebula flares \cite{2012ApJ...746..148C} demonstrated that electrons can be focused into a solid angle of $\Omega_{\rm e}''\simeq 0.1\;{\rm sr}$. However, it would require extreme assumptions ($\mathcal{D}_{\rm VHE}\simeq 200$ or $\Omega_{\rm e}''\simeq 10^{-6}\;{\rm sr}$) to reconcile the above limit with a plausible jet magnetic field strength of $B_{\rm j}'(1\;{\rm pc})\simeq 0.2\;{\rm G}$.

\subsection{Hadronic processes}
\label{sec_rad_hadr}

Alternatively to the leptonic IC scenario, many authors consider processes involving ultra-relativistic protons to explain the gamma-ray emission of blazars \citep[\eg,][]{1993A&A...269...67M,2003APh....18..593M,2009ApJ...703.1168B,2012ApJ...749..119B}. Here, we calculate cooling ratios for the two most popular hadronic interactions: the proton-synchrotron and photo-meson processes.

The proton-synchrotron process is completely analogous to the electron-synchrotron process. In this context, the only relevant difference between protons and electrons is their mass, with the ratio of $m_{\rm p}/m_{\rm e}=1840$. For a fixed synchrotron luminosity and peak energy, the cooling ratio of the synchrotron process scales with the particle mass $m$ as $Q_{\rm SYN}\propto\sigma_{\rm T}\gamma/m\propto m^{-5/2}$, since $\sigma_{\rm T}\propto m^{-2}$ and $\gamma\propto m^{1/2}$. Hence, for proton-synchrotron emission we have
\bea
\gamma_{\rm p,pSYN} &\simeq& 2.6\times 10^{10}B_0''^{-1/2}\left(\frac{\mathcal{D}_{\rm VHE}}{20}\right)^{-1/2}\,,\\
\label{eq_Qpsyn}
Q_{\rm pSYN} &=& \left(\frac{m_{\rm p}}{m_{\rm e}}\right)^{-5/2}Q_{\rm SYN}\simeq\nonumber\\
&\simeq& 3\times 10^{-5}\times B_0''^{3/2}\left(\frac{\mathcal{D}_{\rm VHE}}{20}\right)^{1/2}\,.
\eea

In the photo-meson process, protons interact with soft radiation fields, producing pions of rest energy $m_{\rm\pi} c^2\simeq 140\;{\rm MeV}$ that decay and generate leptonic cascades. In order to create a pion in the interaction with a soft photon of energy $\mathcal{E}_{\rm soft}''$, the proton must have a Lorentz factor satisfying the condition
\be
\gamma_{\rm p,p\gamma}\gtrsim 0.25\left(2+\frac{m_{\rm\pi}}{m_{\rm p}}\right)\frac{m_{\rm\pi}c^2}{\mathcal{E}_{\rm soft}''}\simeq 8\times 10^7\times\left(\frac{\mathcal{E}_{\rm soft}''}{1\;{\rm eV}}\right)^{-1}.
\ee
The typical energy of the created pion is $\mathcal{E}_{\rm\pi}''\simeq \gamma_{\rm p,p\gamma}m_{\rm\pi}c^2$. The effective cross-section for this interaction is $\left<\sigma_{\rm p\gamma}K_{\rm p\gamma}\right>\simeq 7\times 10^{-29}\;{\rm cm^2}$ \citep{1990ApJ...362...38B}. Hence, the cooling ratio can be estimated as
\bea
Q_{\rm p\gamma} &\simeq& \left(\frac{m_{\rm\pi}}{m_{\rm p}}\right)\left<\sigma_{\rm p\gamma}K_{\rm p\gamma}\right>R_{\rm VHE}''n_{\rm soft}''\simeq\nonumber\\
&\simeq& 2.6\times 10^{-15}\times n_{\rm soft,0}''\left(\frac{\mathcal{D}_{\rm VHE}}{20}\right)\,,
\eea
where $n_{\rm soft}''$ is the number density of soft photons. For the external IR radiation of the dusty torus, the co-moving soft photon energy is $\mathcal{E}_{\rm soft}''=\Gamma_{\rm VHE}\mathcal{E}_{\rm IR}$ and thus the proton Lorentz factor must satisfy $\gamma_{\rm p,p\gamma,IR}\gtrsim 1.25\times 10^7(\Gamma_{\rm VHE}/20)^{-1}$. The co-moving number density of soft photons is
\be
n_{\rm soft}'' = \frac{\Gamma_{\rm VHE}u_{\rm IR}}{\mathcal{E}_{\rm IR}} \simeq 4\times 10^9\;{\rm cm^{-3}}\times\left(\frac{\Gamma_{\rm VHE}}{20}\right)\,,
\ee
and hence the cooling ratio is
\be
Q_{\rm p\gamma,IR}\simeq 10^{-5}\times \left(\frac{\mathcal{D}_{\rm VHE}}{20}\right)^2\left(\frac{\mathcal{D}_{\rm VHE}}{\Gamma_{\rm VHE}}\right)^{-1}\,.
\ee

For both processes, we find $Q\ll 1$, even if we assume extremely high values for the Doppler factor and/or the magnetic field strength. We conclude that hadronic processes face serious difficulties in satisfying the energetic requirement for ultra-relativistic protons. This conclusion is in line with a recent review of FSRQ properties by \cite{2009ApJ...704...38S}, and also with the analysis of VHE emission from 3C~279 by \cite{2009ApJ...703.1168B}. In Section \ref{sec_dis_rad_had}, we discuss two specific hadronic scenarios that provide different solutions to the problem of low radiative efficiency.

\subsection{Comparison of radiative processes}
\label{sec_rad_comp}

\begin{figure}
\includegraphics[width=\columnwidth]{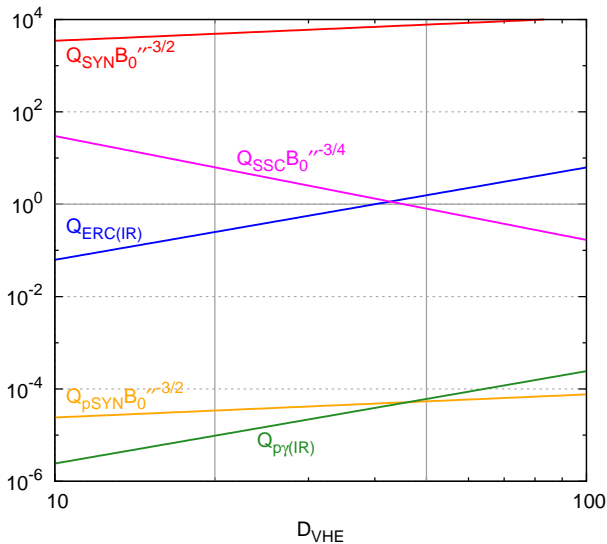}
\caption{Cooling ratios $Q=t_{\rm VHE}''/t_{\rm cool}''$ for all processes considered in Section \ref{sec_rad} as functions of the Doppler factor of the VHE source: inverse Compton scattering off the external IR radiation (ERC(IR); blue) or off the local synchrotron radiation (SSC; magenta), electron synchrotron (SYN; red), proton synchrotron (pSYN; orange), and photo-meson scattering off external IR radiation (${\rm p\gamma(IR)}$; green). Cooling ratios for electron and proton synchrotron processes are multiplied by a factor of $B_0''^{-3/2}$, and that for the SSC process is multiplied by a factor of $B_0''^{-3/4}$. The energetic constraints can be satisfied preferentially by radiatively efficient processes with $Q\gtrsim 1$.}
\label{fig_Qcool}
\end{figure}


The results of this section are summarized in Figure \ref{fig_Qcool}. Hadronic processes are much less efficient than leptonic processes, although for the proton-synchrotron process this conclusion depends on the magnetic field strength. The ERC(IR) process is moderately efficient, but for $B''\gtrsim 0.03\;{\rm G}\times(\mathcal{D}_{\rm VHE}/20)^5$ it will be dominated by the SSC process. The synchrotron process is the most efficient of all for $B''\gtrsim 0.05\;{\rm G}$, assuming typical Doppler factors. However, the strong radiative synchrotron losses limit the electron acceleration process in this scenario to magnetic reconnection sites. Energetic constraints are severe for magnetic reconnection to explain the VHE emission from PKS~1222+216 at the observed luminosity level.

\section{Discussion}
\label{sec_dis}

The flare detected by MAGIC in PKS~1222+216 is an extreme phenomenon that calls for an extreme solution. But it is not our intention to reject the established paradigm of broad-band emission of blazars over longer variability time scales. The entire SED of PKS~1222+216 below 70 GeV has to be explained by a conventional population of electrons with maximum Lorentz factors $\gamma_{\rm e,max}\simeq 10^{3-4}$, producing two spectral components via synchrotron and ERC mechanisms, and most likely filling the entire jet cross section. The evidence for this comes from the relatively long variability time scale of the concurrent HE emission \citep{2011A&A...534A..86T}.

\subsection{Local Energetic Constraint}
\label{sec_dis_lec}

In Section \ref{sec_ene}, we defined the Local Energetic Constraint (LEC) based on an instantaneous energy content within the emitting region. We also showed that a similar result can be derived by considering the rate of energy inflow through the region's boundaries, if the inflow velocity is close to $c$. This is not the case for the magnetic reconnection, where the energy inflow rate is limited by the reconnection rate $\beta_{\rm rec}\lesssim 0.1$. It is unclear whether a flare that satisfies the LEC formulated by Equation (\ref{eq_LEC_uj}) can be sustained at the maximum luminosity for a time longer that $t_{\rm VHE}''$ in any realistic energy dissipation scenario. Nevertheless, this idealized approach to the problem of energetics is very useful in demonstrating that some extreme jet parameters should be adopted. We noted that, independently of the radiative process, we need an extremely high jet power, a very high Doppler factor, very strong jet collimation and/or a highly anisotropic particle beam in the jet co-moving frame.

\subsubsection{Jet power}
\label{sec_dis_lec_pow}

The total jet power for PKS~1222+216 was estimated by \cite{2011ApJ...740...98M}, who used radio flux measurements at $300\;{\rm MHz}$ and obtained $L_{\rm j,M11}\simeq 8\times 10^{44}\;{\rm erg\,s^{-1}}$. This value is not sufficient to satisfy the GEC for the HE flare, since it is smaller than $L_{\rm j,HE}$, given by Equation (\ref{eq_GEC_he}), by a factor of $\simeq 24$. It is also slightly lower than the Poynting flux given by Equation (\ref{eq_LB}). Hence, in Section \ref{sec_rad}, we adopted $L_{\rm j,HE}$ as the preferred value for the total jet power in PKS~1222+216.

We note that the jet power estimated by \cite{2011ApJ...740...98M} may not be directly relevant for the extreme VHE flare. Low-frequency radio emission is strongly self-absorbed at parsec scales and hence it can only probe very extended regions. The variability time scale for this emission is of the order of decades, and thus it can only inform us about the jet power averaged over a very long period of time. If the flare is accompanied by a large increase in the total jet power, the dissipation efficiency $\eta_{\rm diss}$ at parsec scales could be very high, allowing only a small fraction of the initial jet power to reach larger scales. On the other hand, if the jet power increased significantly at constant dissipation efficiency, we expect that variations in the resolved radio structure will follow, most likely including the emergence of a bright superluminal radio spot \citep[\eg,][]{2002A&A...394..851S}. In fact, a short-lived superluminal radio feature was observed with VLBA, with ejection time (defined as the moment of crossing the radio core) estimated at February/March 2010 \citep{2012IJMPS..08..356J}.

The jet power can be conveniently parametrized with the accretion disk luminosity $L_{\rm d}$ \citep[\eg,][]{2008MNRAS.385..283C}. Using the typical radiative efficiency of the accretion disk $\eta_{\rm d}=L_{\rm d}/(\dot{M}_{\rm acc}c^2)\simeq 0.1$, where $\dot{M}_{\rm acc}$ is the mass accretion rate onto the supermassive black hole, we can calculate the jet production efficiency:
\be
\eta_{\rm j} = \frac{L_{\rm j}}{\dot{M}_{\rm acc}c^2} \simeq \left(\frac{L_{\rm j}}{L_{\rm d}}\right)\eta_{\rm d} \simeq 0.04\times\eta_{\rm d,-1}\left(\frac{L_{\rm j}}{L_{\rm j,HE}}\right)\,. 
\ee
This value is relatively low, meaning that the jet power $L_{\rm j,HE}$ adopted in this work is rather modest. \cite{2011MNRAS.418L..79T} demonstrated by numerical simulations that $\eta_{\rm j}\gtrsim 1$ is in principle possible. \cite{2011ApJ...733...19T} obtained $\eta_{\rm j}\simeq 1$, assuming a lower jet Lorentz factor $\Gamma_{\rm j}\simeq 10$, and a much lower accretion disk luminosity $L_{\rm d}\simeq 5\times 10^{45}\;{\rm erg\,s^{-1}}$. For $\eta_{\rm j}\simeq 1$, we would expect the jet power to be as high as $L_{\rm j}\simeq 5\times 10^{47}\;{\rm erg\,s^{-1}}$. This is still lower, by almost 3 orders of magnitude, than the jet power required by the LEC for the VHE flare (see Equation \ref{eq_LEC_Lj}).

\subsubsection{Doppler factor}
\label{sec_dis_lec_dop}

A Lorentz factor of $\Gamma_{\rm VHE}=50$ and a Doppler factor of $\mathcal{D}_{\rm VHE}=75$ were proposed by \cite{2011A&A...534A..86T} in their interpretation of the VHE flare in PKS~1222+216. Similar relativistic boosts were also suggested for previous cases of fast TeV flares in other blazars \citep{2008MNRAS.384L..19B}. Such values for the bulk Lorentz factor are much higher than those inferred from the apparent superluminal motion of individual pc-scale radio features, which typically fall into the $10-20$ range, with exceptional cases up to $\simeq 40$ \citep[\eg][]{2009A&A...494..527H}. Several solutions to this discrepancy have been proposed, including radiative deceleration \citep{2007ApJ...671L..29L}, the photon breeding mechanism \citep{2008MNRAS.383.1695S}, stratified jet emission \citep{2008MNRAS.390L..73B}, minijets \citep{2009MNRAS.395L..29G,2011MNRAS.413..333N} and non-steady magnetic acceleration \citep{2010ApJ...722..197L}. The minijets model postulates that localized relativistic outflows are produced in the jet co-moving frame perpendicularly to the jet bulk motion and powered by Petschek-type magnetic reconnection. However, in order for the minijets to have significant Lorentz factors, the reconnecting plasma must be strongly magnetized, with $\sigma\gg 1$. In the case of PKS~1222+216, where VHE emission should be produced at $r>r_{\rm min}\simeq 0.5\;{\rm pc}$, the jet plasma should already be matter-dominated. For this reason, \cite{2011A&A...534A..86T} judged that minijets cannot operate at such large scales.

\cite{2012MNRAS.420..604N} proposed in the context of fast VHE variability of blazars that localized relativistic outflows are driven by turbulence, rather than magnetic reconnection. Multi-zone turbulent models were proposed to explain the broad-band emission of blazars that do not show correlated variability across different bands \citep[\eg,][]{2012arXiv1201.5402M}. However, the origin of such a relativistic turbulence is not specified and it is not clear if such models are energetically viable. In the case of PKS~1222+216, the main concern is whether these models provide a solution to the extremely high energy density problem.

The main motivation for introducing high Doppler factors of $\mathcal{D}_{\rm VHE}\simeq 50$ in previous cases of fast VHE flares was to avoid the internal absorption of the VHE emission originating in a very compact region \citep{2008MNRAS.384L..19B}. In the case of PKS~1222+216, we found insignificant internal absorption for $\mathcal{D}_{\rm VHE}\simeq 20$. The reasons for this apparent inconsistency are the following: in this work the observed photon energy is $\mathcal{E}_{\rm VHE,obs}\simeq 100\;{\rm GeV}$ instead of $1\;{\rm TeV}$; the soft radiation luminosity is $L_{\rm soft}\simeq 10^{45}\;{\rm erg\,s^{-1}}$ instead of $10^{46}\;{\rm erg\,s^{-1}}$; the observed variability time scale is $t_{\rm VHE,obs}\simeq 10\;{\rm min}$ instead of $5\;{\rm min}$; and we took into account the redshift and the 3.6 factor in Equation (\ref{eq_Esoft_vhe}). Different approximations lead to a difference by factor $\simeq 21$ between our Equation (\ref{eq_tauVHE}) and their Equation (16). Even if we drop the 3.6 factor and consider the $\mathcal{E}_{\rm VHE,max,obs}=400\;{\rm GeV}$ photons, we obtain $\mathcal{E}_{\rm soft,VHE,obs}\simeq 125\;{\rm eV}\times(\mathcal{D}_{\rm VHE}/20)^2$ and $\tau_{\rm \gamma\gamma,VHE}\simeq 0.6L_{\rm soft,45}(\mathcal{D}_{\rm VHE}/20)^{-6}$. Hence, the VHE emission in PKS~1222+216 could be produced in a region propagating with a typical Lorentz factor.

\subsubsection{Reconfinement shock nozzle}
\label{sec_dis_lec_rec}

The radius-to-distance ratio of the VHE emitting region, if located at the broad-line photosphere $r_{\rm min}$, is $\simeq 1.7\times 10^{-4}$ (see Equation \ref{eq_theta_vhe}). \cite{2011ApJ...730L...8A} and \cite{2011A&A...534A..86T} proposed that such a small value can be explained by a reconfinement shock that results from the interaction between the jet and the pressurized external medium. Reconfinement shocks have been proposed as the main dissipation mechanism operating at the scales of a few pc in blazars \citep{2008ApJ...675...71S,2009MNRAS.392.1205N}. According to the study by \cite{2009ApJ...699.1274B}, reconfinement shocks can focus relativistic jets very efficiently, but this effect depends strongly on the radiative efficiency of the downstream flow. The radius-to-distance ratio inferred in the case of PKS~1222+216 is even smaller than the value of $\simeq 5\times 10^{-4}$ required by the hypothesis that the 2006 TeV flare in M87 originated in the quasi-stationary knot HST-1 \citep{2006Sci...314.1424A,2006MNRAS.370..981S}. \cite{2009ApJ...699.1274B} considered that hypothesis unrealistic, noting the small radiative efficiency of the M87 jet. For radiative efficiency of $30\%$ they obtained a radius-to-distance ratio of $\simeq 3\times 10^{-3}$. A somewhat larger value of $\simeq 6\times 10^{-3}$ has been obtained for a feature resembling HST-1 in relativistic MHD numerical simulations designed specifically for the case of M87 \citep{2009ApJ...695..503G}.

These opening angles are more than order of magnitude larger than that required in the case of PKS~1222+216. Moreover, the extreme jet focusing mechanism via reconfinement shocks, as proposed by \cite{2009ApJ...699.1274B}, depends critically on the assumption of perfect jet axisymmetry. The jet of PKS~1222+216 is unlikely to be axisymmetric, since individual radio elements were observed to propagate along a curved trajectory \citep{2009ApJ...706.1253H}. Hence, the interpretation of the VHE source compactness as due to a reconfinement nozzle is unlikely. However, numerical simulations \citep[\eg,][]{2009ApJ...695..503G} indicate that reconfinement can result in reducing the jet radius by one order of magnitude, as compared to the conical jet, helping to relax the LEC.

\subsubsection{Jet substructure}
\label{sec_dis_sub}

The VHE flare could be associated with the appearance of a very compact and energetic substructure within the jet \citep[\eg,][]{2008MNRAS.386L..28G}. This allows one to abandon the LEC and require only that the GEC be satisfied. Should the radius of this substructure correspond to the observed variability time scale of the VHE flare, a substantial fraction of the total jet power, $(L_{\rm j,VHE,GEC}/L_{\rm j,HE})\simeq 15\%\times(\Omega_{\rm e}''/4\pi)$, would be carried through a tiny fraction of the jet cross section, $(R_{\rm VHE}''/R_{\rm j}')^2\simeq 3\times 10^{-4}\times(\mathcal{D}_{\rm VHE}/20)^4$. This extreme requirement could be relaxed somewhat by increasing the Doppler factor $\mathcal{D}_{\rm VHE}$ or by considering an anisotropic co-moving distribution of emitting particles, with $\Omega_{\rm e}''\lesssim 1\;{\rm sr}$, as expected in the extreme particle acceleration scenario. However, in order to maintain a sharply higher energy density in the substructure, compared to the broad jet, a very efficient self-collimating process must operate. This can be provided by the pinch mechanism, if the structure is highly magnetized. The high magnetization would imply a very strong magnetic field that would be potentially sufficient to power the VHE flare and associated extreme particle acceleration via magnetic reconnection (see Section \ref{sec_rad_syn_acc}). High-magnetization regions could be associated with jet cores expected to arise along the axis of the light cylinder \citep{2010PhyU...53.1199B}. Another phenomenon enabled by a high-density magnetized structure propagating through a low-density jet medium is the magnetic rocket effect, in which a thin shell is accelerated to large Lorentz factors, while its magnetization decreases to $\sigma\ll 1$ \citep{2011MNRAS.411.1323G}. This allows, at least locally, very efficient conversion of Poynting flux to kinetic power. The energy of such an ultra-relativistic but weakly magnetized shell could be efficiently dissipated via fast magnetosonic waves \citep{2012MNRAS.422..326K}. Such a scenario has been already applied to the fast TeV flares in blazars \citep{2010ApJ...722..197L}.

\subsubsection{Kinetic beaming}
\label{sec_dis_kin_beam}

Recent Particle-in Cell (PIC) simulations by \cite{2012arXiv1205.3210C} show that the observed variability time scale of emission associated with dissipation via relativistic magnetic reconnection can be one order of magnitude shorter than the light-crossing time scale of the reconnection layer. This is possible when the highest-energy particles form very coherent beams that are either short-lived or sweep across certain lines of sight and produce short pulses of non-thermal radiation. It is crucial that acceleration sites for these particles reprocess a significant fraction of the reconnecting magnetic energy and hence this process can be very efficient in converting the magnetic energy into high-energy radiation.

If a similar mechanism operates in the pc-scale jet of PKS~1222+216, one can relax the causality constraint on the size of the emitting region and adopt $R_{\rm VHE}''\simeq 10\,ct_{\rm VHE}''$. This reduces the requirement for the energy density of the emitting region by 2 orders of magnitude. In addition, the kinetic beaming of the most energetic particles can reduce the solid angle occupied by their momenta to $(\Omega_{\rm e}''/4\pi)\sim 0.01$. These two effects combined can reduce the local energy density to the levels typical for pc-scale jets. Moreover, the results of \cite{2012arXiv1205.3210C} were obtained for very moderate electron energies and they do not rely on the extreme particle acceleration mechanism described in Section \ref{sec_rad_syn_acc}. Hence, in this scenario the MAGIC emission could be produced by the SSC process for typical Doppler factors or by the ERC(IR) process for large Doppler factors.

\subsection{Radiative processes}
\label{sec_dis_rad}

\subsubsection{Synchrotron radiation}
\label{sec_dis_rad_syn}

Using electron synchrotron radiation to interpret the VHE emission from a blazar would certainly be an unconventional scenario. Although the synchrotron mechanism has by far the highest cooling ratio of all processes considered in Section \ref{sec_rad} (for magnetic field strengths of order $B''\simeq 0.2\;{\rm G}$, typical for parsec-scale blazar jets, see Section \ref{sec_obs_mag}), this feature cannot be exploited because of the difficulty in accelerating electrons to the required Lorentz factors $\gamma_{\rm e,SYN}\simeq 10^9$. As we showed in Section \ref{sec_rad_syn_acc}, in order to avoid severe radiative losses, the acceleration should proceed in an environment where the electric-to-magnetic field strength ratio is $E''/B_\perp''\simeq 26\times (\mathcal{D}_{\rm VHE}/20)^{-1}$, which can be satisfied only very deep within a magnetic reconnection layer. This large value by itself poses a serious problem, since it is hard to imagine the focusing mechanism working so perfectly under realistic conditions \citep{2012ApJ...746..148C}. Increasing the source Doppler factor can relax this requirement, but probably not by a large enough factor. Moreover, the typical magnetic field strengths for parsec-scale jets are insufficient by almost 3 orders of magnitude to explain the energetics of the VHE flare. These two problems make the synchrotron scenario for the VHE flare in PKS~1222+216 implausible.

\subsubsection{Synchrotron self-Compton radiation}
\label{sec_dis_rad_ssc}

Synchrotron self-Compton (SSC) radiation is generally thought to dominate the gamma-ray emission in BL Lac objects, but not in FSRQs \citep[\eg,][]{2009ApJ...704...38S}. In the case of the MAGIC flare in PKS~1222+216, the exceptional requirement for a very compact emitting region implies a very high energy density of the local synchrotron radiation, even if the apparent luminosity of this component is relatively low. SSC was favored by \cite{2009ApJ...703.1168B} as the mechanism of VHE emission in another FSRQ, 3C~279. In cases when the SSC luminosity dominates the synchrotron luminosity, the SSC cooling proceeds in a non-linear regime \citep{2012MNRAS.420...84Z}. This allows a somewhat shorter cooling time scale compared to a simplified linear SSC model, which could make the SSC scenario even more favorable.

\subsubsection{Hadronic processes}
\label{sec_dis_rad_had}

\cite{2012ApJ...749..119B} interpreted the rapid VHE flares in BL Lacs in terms of proton synchrotron emission associated with the interaction between the jet and a red giant star. In order to make the proton synchrotron mechanism efficient, they adopted magnetic field strengths of the order of $\simeq 100\;{\rm G}$ and located the emission site at the distance of $\simeq 0.003\;{\rm pc}$ from the supermassive black hole. However, in the case of PKS~1222+216, the VHE emission is constrained to be produced at the distance larger than $r_{\rm min}\sim 0.5\;{\rm pc}$ (see Section \ref{sec_obs_opac}). Such a high magnetic field strength is not expected to occur at this scale. Even if it could be realized there, it would have to be contained in a very narrow beam, otherwise the required total jet power would be too high (see the discussion of the LEC in Section \ref{sec_ene}).

\cite{2012arXiv1203.6544D} proposed a model of the MAGIC flare in PKS~1222+216 based on the assumption that its inner jet is a source of Ultra-High-Energy Cosmic Rays (UHECRs). In this model, ultra-relativistic neutrons form a tightly collimated beam that remains coherent up to the parsec scale, where they interact with the infrared radiation from the dusty torus and deposit their energy in lepto-mesonic cascades. In order to preserve the short variability time scale, relatively low values of the jet magnetic field are required, $\lesssim 0.01\;{\rm G}$. The gamma rays are produced mainly by the synchrotron process. In order to satisfy the energetic requirements, an extremely large Doppler factor is required, $\mathcal{D}_{\rm VHE}\sim 100$. Even with such a high value of the Doppler factor, the conversion of the energy of ultra-relativistic hadrons proceeds over a time scale much longer than the observed variability time scale. This is possible only in a strongly anisotropic particle beam.

\section{Summary}
\label{sec_sum}

We studied theoretical implications of a VHE flare of flux-doubling time scale of $t_{\rm VHE,obs}\simeq 10\;{\rm min}$ observed by MAGIC at photon energies between 70~GeV and 400~GeV in the FSRQ-class blazar PKS~1222+216. Assuming negligible absorption of the VHE emission by the BLR radiation, we require that its source be located at the distance of at least $r_{\rm min}\simeq 0.5\;{\rm pc}$ from the supermassive black hole. Contrary to the cases of rapid VHE flares in PKS~2155-304 and Mrk~501, a very high Doppler factor is not implied by the internal gamma-ray opacity argument.

We analyzed energetic constraints that have to be satisfied by any candidate radiative process behind the VHE emission. Assuming the maximum radiative efficiency of $\eta_{\rm rad}\simeq 1$, and a moderate dissipation efficiency of $\eta_{\rm diss,VHE}\simeq 0.1$, the energy density of the VHE emitting particles should not exceed the local energy density of the jet, thus estimated at $u_{\rm j,VHE}''\simeq 1200\;{\rm erg\,cm^{-3}}\times(\Omega_{\rm e}''/4\pi)(\mathcal{D}_{\rm VHE}/20)^{-6}$. The power carried by the VHE emitting particles is $L_{\rm j,VHE,GEC}\simeq 2.8\times 10^{45}\;{\rm erg\,s^{-1}}\times (\Omega_{\rm e}''/4\pi)(\mathcal{D}_{\rm VHE}/20)^{-2}$, while the most likely total jet power can be estimated from the HE luminosity, $L_{\rm j,HE}\simeq 1.9\times 10^{46}\;{\rm erg\,s^{-1}}\times (\mathcal{D}_{\rm j}/20)^{-2}$. Hence, a substantial fraction of the total jet power, $(L_{\rm j,VHE,GEC}/L_{\rm j,HE})\simeq 15\%\times(\Omega_{\rm e}''/4\pi)$, needs to be carried through a tiny fraction of the jet cross section, $(R_{\rm VHE}''/R_{\rm j}')^2\simeq 3\times 10^{-4}\times(\mathcal{D}_{\rm VHE}/20)^2(\mathcal{D}_{\rm j}/20)^2$. This puzzle of an extreme energy concentration at the parsec scale challenges every model of the structure of relativistic jets in blazars. While the energy density constraint could be relaxed by increasing the total jet power, by increasing the Doppler factor or by assuming a very strong jet collimation, the required parameter values are too extreme for AGN jets.

Out of several possibilities discussed in Section \ref{sec_dis}, we favor the kinetic beaming of particles accelerated via magnetic reconnection \citep{2012arXiv1205.3210C} as the most plausible scenario for producing extremely fast VHE variability of PKS~1222+216 (see Section \ref{sec_dis_kin_beam}). This scenario allows one to reduce the required energy density within the emitting region by: 1) relaxing the causality constraint on the emitting region size $R_{\rm VHE}''\sim 10\,ct_{\rm VHE}''$ due to the energy clustering and swinging of the particle beams, and 2) focusing the accelerated particles into a narrow solid angle $\Omega_{\rm e}''\sim 0.1$. The VHE emission could be due to the SSC process for a moderate Doppler factor $\mathcal{D}_{\rm VHE}\simeq 20$, or the ERC(IR) process for a large Doppler factor $\mathcal{D}_{\rm VHE}\simeq 50$.

\section*{Acknowledgments}
We thank the reviewers for their very helpful comments that significantly improved early versions of the manuscript. This work was partly supported by the NSF grant AST-0907872, the NASA ATP grant NNX09AG02G, and the Polish NCN grant DEC-2011/01/B/ST9/04845.

\end{document}